\RequirePackage{fix-cm}
\documentclass[twocolumn]{svjour3}          
\smartqed  
\usepackage{makecell}
\usepackage{graphicx}
\usepackage{hyperref}
\usepackage{amsmath,amssymb,amsfonts}
\usepackage{algorithmicx}
\usepackage{algorithm,algpseudocode}

\usepackage{physics}
\usepackage{multirow}
\usepackage{booktabs}
\usepackage{xcolor}
\usepackage{textcomp}
\usepackage{threeparttable}
\usepackage{pgfplotstable}
\usepackage{caption}
\usepackage{subcaption}

\usepackage{bbding}

\usepackage[numbers, sort&compress]{natbib}

\pgfplotsset{compat=1.16}
\definecolor{darkred}{rgb}{0.55, 0.0, 0.0}
\definecolor{darkgoldenrod}{rgb}{0.72, 0.53, 0.04}
\definecolor{darkkhaki}{rgb}{0.74, 0.72, 0.42}
\definecolor{navyblue}{rgb}{0.0, 0.0, 0.5}
\definecolor{moonstoneblue}{rgb}{0.45, 0.66, 0.76}
\definecolor{ashgrey}{rgb}{0.7, 0.75, 0.71}
\definecolor{blond}{rgb}{0.98, 0.94, 0.75}
\definecolor{burntumber}{rgb}{0.54, 0.2, 0.14}
\definecolor{darkcyan}{rgb}{0.0, 0.55, 0.55}

\usepackage{pgfplots}
    \usepgfplotslibrary{colorbrewer}
    \pgfplotsset{compat=1.17}
\begin{document}

\title{An investigation of IBM Quantum Computing device performance on Combinatorial Optimisation Problems
}


\author{  Maxine T. Khumalo\and Hazel A. Chieza \and Krupa Prag \and Matthew Woolway}

\authorrunning{  M. T. Khumalo \and H. A. Chieza \and K. Prag \and M. Woolway} 

\institute{Maxine T. Khumalo \at
              University of the Witwatersrand \\
              School of Computer Science and Applied Mathematics \\
              \email{1604282@students.wits.ac.za}           
           \and
           Hazel A. Chieza \at
              University of the Witwatersrand \\
              School of Computer Science and Applied Mathematics \\
              \email{1609247@students.wits.ac.za}   
           \and
           \Envelope\ Krupa Prag \at
              University of the Witwatersrand \\
              School of Computer Science and Applied Mathematics \\
              \email{krupa.prag@wits.ac.za}   
           \and 
           Matthew Woolway \at
              University of Johannesburg \\
              Faculty of Engineering and the Built Environment \\
              \email{mjwoolway@uj.ac.za}
}

\date{Received: date / Accepted: date}

\maketitle

\begin{abstract}


The intractability of deterministic solutions in solving $\mathcal{NP}$-Hard Combinatorial Optimisation Problems (COP) is well reported in the literature. One mechanism for overcoming this difficulty has been the use of efficient COP non-deterministic approaches. However, with the advent of quantum technology, these modern devices' potential to overcome this tractability limitation requires exploration. This paper juxtaposes classical and quantum optimisation algorithms' performance to solve two common COP, namely the Travelling Salesman Problem (TSP) and the Quadratic Assignment Problem (QAP). Two accepted classical optimisation methods, Branch and Bound (BNB) and Simulated Annealing (SA), are compared to two quantum optimisation methods, Variational Quantum Eigensolver (VQE) algorithm and Quantum Approximate Optimisation Algorithm (QAOA). These algorithms are respectively executed on both classical devices and IBM's suite of Noisy Intermediate-Scale Quantum (NISQ) devices. We have encoded the COP problems for the respective technologies and algorithms and provided the computational encodings for the NISQ devices. Our experimental results show that current classical devices significantly outperform the presently available NISQ devices, which both agree with and extend on those findings reported in the literature. Further, we introduce additional performance metrics to better compare the two approaches with respect to computational time, feasibility and solution quality. Our results show that the VQE performs better than QAOA with respect to these metrics, and we infer that this is due to the increased number of operations required. Additionally, we investigate the impact of a new set of basis gates on the quantum optimisation techniques and show they yield no notable improvement on obtained results. Finally, we highlight the present shortcomings of state-of-the-art NISQ IBM quantum devices and argue for continued future work on improving evolving devices.



\end{abstract}

\section{Introduction}
\label{intro}

Two important examples of $\mathcal{NP}$-Hard Combinatorial Optimisation Problems (COP) are the Travelling Salesman Problem (TSP) and Quadratic Assignment Problem (QAP). The TSP requires finding the shortest route to visit all enlisted locations \cite{TSPformulation}, while the QAP aims to find the minimum cost of allocating facilities to locations \cite{10.2307/1907742}. Both TSP and QAP can be translated to solve numerous industry challenges. Common TSP real-world applications found in the literature include X-ray crystallography \cite{BLAND1989125}, computer wiring \cite{lenstra1975some}, and various routing and scheduling problems \cite{dantzig1959truck,laporte1992traveling}. Similarly, some use cases of the QAP formulation include energy system problems \cite{ajagekar2019quantum}, backboard wiring \cite{burkard1998quadratic}, typewriter keyboard design \cite{burkard1977entwurf}, the hospital layout challenge \cite{elshafei1977hospital}, and campus building arrangement \cite{dickey1972campus}.

Solutions to these combinatorial problems have been extensively studied using various optimisation techniques implemented on classical devices \cite{laporte1992traveling,burkard1998quadratic}. For example, recent classical literature points to advancement in selected TSP and QAP applications; such as Particle Swarm Optimisation (PSO) for the generation design of non-trivial flat-foldable origami tessellations with degree-4 vertices \cite{chen2021particle}; Ant Colony Optimisation (ACO) for optimal skeletal structures and novel form-finding of tensegrity structures \cite{chen2012novel}; graph theory and mixed-integer linear programming for assigning mountain-valley fold lines of flat-foldable origami patterns based \cite{chen2020assigning}. However, recent advances in Quantum technology have led to the investigation of these quantum devices' performance when applied to various COP instances. Specifically, through the use of Noisy Intermediate-Scale Quantum (NISQ) technology, preliminary findings for both the TSP \cite{warren2017small,srinivasan2018efficient} and QAP \cite{ajagekar2019quantum} have been reported. Furthermore, a detailed and comparative investigation into the performance of the Variational Quantum Eigensolver (VQE) algorithm run on these NISQ devices against classical devices \cite{chieza2020computational} showed the classical optimisation techniques significantly outperform the suite of IBM devices utilised. 

In this paper, we extend the state-of-the-art benchmarks reported in \cite{chieza2020computational} by including the newest and largest current IBM devices available. We further introduce preliminary findings for the Quantum Approximate Optimisation Algorithm (QAOA) on amenable tractable problem instances. Additionally, we present a feasibility metric and a new metric to determine the spectrum of the feasibility of quantum algorithms on the ensemble of IBM devices. Finally, we investigate the computational performance improvements by utilising the \textit{conditional reset} feature on the IBM systems over those used in \cite{chieza2020computational}.

The remainder of the paper is organised as follows. Section~\ref{sec:background} presents the complexity classification and the formulation of the TSP and QAP. Section~\ref{sec:solution_techniques} discusses the techniques and algorithms used for both the classical and Quantum devices. Section~\ref{sec:methodology} details the experimental procedure and settings used in conducting the experiments. Results and analysis are presented in Section~\ref{sec:results}, with conclusions drawn in Section~\ref{sec:conclusion}.

\section{Background}\label{sec:background}

The TSP has been extensively studied and reviewed. Seminal research of the TSP occurred in 1959 with the Dantzig-Fulkerson-Johnson (DFJ) \cite{TSPformulation} formulation, detailing how the TSP problem instances of up to 52 locations could be solved using an exact algorithm. The DFJ formulates the TSP as an Integer Linear programming (ILP) and then utilises an exact cutting-plane algorithm to solve the LP relaxation of the ILP. The TSP forms the basis of many complex COP that model the intricacies of real-world scenarios. The complexity class of the TSP is $\mathcal{NP}$-Hard \cite{Korte2012}, which makes it intractable to exactly solve the problem using deterministic algorithms as the problem size scales. Therefore, research developments of the TSP have turned the focus on non-deterministic heuristic and meta-heuristic methods. One of the most effective heuristics in the literature is the Lin-Kerninghan (LK) algorithm \cite{lin1973effective}, which utilises an adaptive algorithm that consists of swapping the edges of sub-tours to create a new tour with an overall minimal distance. Other notable heuristics and meta-heuristic algorithms which have been applied to the TSP include; the Simulated Annealing (SA) algorithm \cite{kirkpatrick1983optimization}, Gene Expression Programming \cite{ferreira2001gene}, and the Ant Colony Optimisation algorithm \cite{dorigo2006ant}. These algorithms perform well in the literature, and ongoing research is often an extension, adaptation or combination of these fundamental and well-known algorithms \cite{zhou2019traveling}.

The QAP was first formulated by Koopmans and Beckmann (1957)~\cite{10.2307/1907742}. It found importance due to the added relative cost of assignment with a quadratic objective function to solve assignment class problems, a phenomenon prevalent in industry \cite{10.2307/1907742}. Practically, the QAP is primarily solved via heuristics as similar to the TSP; the problem becomes intractable as the problem instances scale in size. The \href{http://anjos.mgi.polymtl.ca/qaplib/}{QAPLIB} documents a set of QAP benchmark problems and their best solutions as reported in the literature. \href{http://anjos.mgi.polymtl.ca/qaplib/}{QAPLIB} records the following methods as most effective on large instances of the QAP: Branch and Bound \cite{lawler1963quadratic}, Tabu Search \cite{taillard1991robust} and Ant Colony Optimisation \cite{dorigo2006ant,stutzle1997max}.

Using quantum computing methods to solve the TSP and the QAP begins with Computational Complexity Theory. As previously mentioned, the TSP and the QAP are both $\mathcal{NP}$-hard problems \cite{Korte2012,sahni1976p}, which means that as the problem scales in size, it becomes difficult for a classical computer to find an optimal solution in polynomial time \cite{burkard1998quadratic}. Quantum Complexity Theory describes the computational complexity of quantum computers. Quantum Complexity Theory is necessary because there are problems too complex to solve with classical computing methods that quantum computers can theoretically solve \cite{bernstein1997quantum} and these problems computational complexity requires description. Between a classical probabilistic Turing machine and a Quantum Turing Machine (QTM), a polynomial-time QTM is more powerful in certain search algorithms \cite{bennett1997strengths,bernstein1997quantum,deutsch1985quantum}. A basic search algorithm for optimisation problems has complexity $\mathcal{O}(2^n)$ on a classical computer while a quantum search algorithm for an $\mathcal{NP}$-Complete problem reduces this to complexity $\mathcal{O}(\sqrt{n})$ \cite{grover1997quantum,montanaro2016quantum,ronnow2014defining}. Therefore, there is reason to believe that there may be a speed-up from quantum computing when applied to optimisation problems.

\subsection{Mathematical Model}

\subsubsection{Travelling Salesman Problem}
The TSP has multiple mathematical formulations \cite{Korte2012}, however, the DFJ formulation \cite{TSPformulation} has the ability to form a strong LP relaxation which makes it the preferred formulation of the TSP. This formulation is given as follows:
\begin{equation}\label{eq1}
\min \sum_{i = 1}^{n} \sum_{j=1}^{n} d_{ij}x_{ij}
\end{equation}
Subject to:
\begin{equation}\label{eq2}
\sum_{j = 1}^{n}x_{ij} = 1 ,  \hspace{0.5cm}i = 1,...,n,
\end{equation}
\begin{equation}\label{eq3}
\sum_{i = 1}^{n}x_{ij} = 1 , \hspace{0.5cm} j = 1,...,n, 
\end{equation}
\begin{equation}\label{eq4}
\sum_{i,j \in S}x_{ij} \leq \lvert S \rvert - 1,
\end{equation}
\begin{equation}\label{eq5}
S \subset{V} , 2 \leq \lvert S \rvert \leq n - 2, 
\end{equation}
\vspace{-0.35cm}
\begin{equation}\label{eq6}
 x_{ij} \in \{0,1\} ,    
\end{equation}
\vspace{-0.35cm}
\begin{equation}\label{eq7}
i,j = 1,...,n   \qquad i \neq j.   
\end{equation}

Equation~\eqref{eq1} is the objective function that seeks to minimise the total travelled distance, subject to constraints~(\ref{eq2}-\ref{eq7}). The distance between city $i$ and city $j$ can be denoted in the form of a distance matrix by $d_{ij}$. The binary decision variable, $x_{ij}$, indicates whether a path between cities $i$ and $j$ exists.

Constraint~(\ref{eq2}) and constraint~(\ref{eq3}) prevent any city from being visited more than once. The sub-tour elimination constraints~(\ref{eq4}) and constraint~(\ref{eq5}) ensure that all cities of a TSP problem are visited, where $S$ is a subset of the $n$ cities. Because the sub-tour-elimination constraints' variables grow exponentially, it is intractable to directly solve the DFJ formulation. For every value of $n$, there is $2^{n} -2n -2$ sub-tour-elimination constraints and $n(n-1)$ binary decision variables. 

\subsubsection{Quadratic Assignment Problem}

To mathematically model the QAP, weights are defined for the cost associated with moving goods between location $k$ and $l$ ($b_{kl}$) and factories $k$ and $l$ ($c_{kl}$) \cite{10.2307/1907742}. The cost function is then calculated by finding the product of the associating costs of allocated factories to locations. This allocation is denoted by a permutation function $\pi$ that records which factories are allocated to which locations. The elements in $\pi$ are the factory number allocated to the location represented by its position in the vector. The QAP objective function is:
\begin{equation}
    \mbox{min} \quad \sum_{k=1}^{n} \sum_{l=1}^{n} b_{kl}c_{\pi(k),\pi(l)},
    \label{eq:prob}
\end{equation}
where $b$ and $c$ are matrices of dimensions $n \times n$ \cite{burkard1998quadratic}. The full form of the objective function is:
\begin{equation}
\sum_{k=1}^{n} \sum_{l=1}^{n} b_{kl}c_{\pi(k),\pi(l)} + \sum_{k = 0}^{n} a_{k,\pi(k)},
\end{equation}
where the value $a_{kl}$ represents the cost of allocating factory $k$ to factory $l$ is known as the ``cost of assignment". This additional cost term is omitted in the benchmark formulations for ease of calculation \cite{burkard1998quadratic}.

The QAP is a quadratic optimisation problem because of its quadratic objective function. The QAP is also a bijection of a finite set $\mathcal{N}$ because unique position values from set $\mathcal{N}$ are allocated as elements of $\pi$ \cite{burkard1998quadratic}. This characteristic means that the QAP can have a ``0-1" integer optimisation problem formulation \cite{burkard1998quadratic,lawler1963quadratic}. This ``0-1" integer formulation is the version used to solve the QAP with quantum computing.

\subsubsection{Quantum formulations}

To apply quantum algorithms to find solutions of COP problems using a quantum devices, the problems’ objective function first needs to be mapped to a Hamiltonian form \cite{QiskitTextbook:2020}. The respective Hamiltonian of Ising model formulation mappings of the TSP and QAP, described in Section~\ref{sec:background}, are presented in this section.  

The objective and constraints of the TSP can be described by:
\begin{eqnarray}\label{eqn:TSP_quantumFormulation}
&&\sum_{i,j}d_{ij}\sum_{p}x_{ip}x_{jp+1}\nonumber \\ &+& A\sum_{p}\left(1-\sum_{i}x_{ip}\right)^{2} +A\sum_{i}\left(1-\sum_{p}x_{ip}\right)^{2}.
\end{eqnarray}
Similarly, the QAP can be described by:
\begin{eqnarray}\label{eqn:QAP_quantumFormulation}
      &&\sum_{m=1}^{n}\sum_{u=1}^{n}\sum_{k=1}^{n}\sum_{l=1}^{n} C_{lk}T_{um}x_{ul}x_{mk} \nonumber \\
      &+&B\sum_{l}\left(1-\sum_{u}x_{ul}\right)^{2}+ B\sum_{u}\left(1-\sum_{l}x_{ul}\right)^{2},
\end{eqnarray}
by assigning
$$
x_{ip} \rightarrow (1 - Z_{ip})/2,
$$ 
and
$$
x_{ul} \rightarrow (1 - Z_{ul})/2,
$$ 
where $Z_{ij}$ is a Pauli operator that maps the TSP to a Hamiltonian, and $Z_{ul}$ is a Pauli operator that maps the QAP to a Hamiltonian  \cite{QiskitTextbook:2020,lucas2014ising,ajagekar2019quantum}. $A$ and $B$ are free parameters that satisfy the constraints. The VQE and QAOA algorithms are used to find the ground state of the respective Hamiltonian systems for the TSP and QAP, described by equation~(\ref{eqn:TSP_quantumFormulation}) and equation~(\ref{eqn:QAP_quantumFormulation}). The ground state is where the expectation value of the energy state is the lowest. Both quantum algorithms, VQE and QAOA, applied to the TSP and QAP use variational methods which consist of selecting a variational form. This method depends on finding one or more parameters to find approximations to the ground state. Simply, this describes an ansatz in the form of a parametrised circuit. The accuracy of solutions depends on the variational form \cite{moll2018quantum}. The depth of the variational form is the exponential to the number of qubits, hence the parameters of a variational form exponentially increase with the number of qubits. Due to the limited number of qubits currently available on NISQ devices, the size of the problem instances considered is restricted \cite{Rattew2019ADN}. 

The VQE algorithm was proposed by \cite{peruzzo2014variational} to approximate the ground state energy of a molecule \cite{QiskitTextbook:2020} and early applications of VQE to chemistry problems were predominantly studied. Recently, the literature on the VQE applications has been extended to solve COP \cite{moll2018quantum,verteletskyi2020measurement}. Preliminary results using NISQ technology to find solutions to the QAP are shown in \cite{ajagekar2019quantum}. The VQE algorithm’s performance is benchmarked against classical algorithms, SA and BNB, for the TSP and QAP in \cite{chieza2020computational}. The findings in \cite{chieza2020computational} show that the classical algorithms SA and BNB outperform the VQE algorithm with reference to both solution quality and computational time. This work aims to expand on \cite{chieza2020computational}, by introducing additional metrics to measure solution quality and comparing the results to the COPs obtained using the VQE algorithm to an additional quantum algorithm, QAOA. Recent developments on the QAOA are outlined in \cite{zhou2020quantum}, and applications of the QAOA to solve the Maxcut, a prominent COP, have been investigated \cite{crooks2018performance,farhi2017quantum,harrigan2021quantum}. The literature on the QAOA includes studies to improve the QAOA using graph theory to optimise the solving of the  Hamiltonian \cite{verteletskyi2020measurement}, and the $p$-value approach \cite{farhi2020quantum}.

\section{Solution Techniques}\label{sec:solution_techniques}

This section discusses classical and quantum techniques used to obtain solutions to the TSP and QAP.

\subsection{Branch and Bound}\label{sec:BNB}

The Branch and Bound (BNB) algorithm is an exact algorithm that reduces the computational runtime of obtaining a solution as it indirectly computes all possible combinations of a given COP \cite{MORRISON201679}. The BNB first solved a discrete $\mathcal{NP}$-hard optimisation problem \cite{land2010automatic}, namely the British Petroleum TSP problem. The BNB represents optimal and sub-optimal solutions in the form of a tree data structure constructed with multiple levels of artificial nodes. Each node represents a calculated distance for the TSP \cite{gavett1966optimal}, and an allocation permutation vector $\pi$ and product of distance and flow for the QAP \cite{clausen1999branch}. At each level, the node with the lowest value is further explored and broken into sub-problems. This iterative process is terminated when a complete tour is formed for the TSP, and a suitable permutation matrix is formed for the QAP. Nodes are selected to explore using the best-first strategy, which entails comparing the rolling cost function values between levels to explore further which nodes to explore. The efficiency of the BNB algorithm depends on an appropriate estimation of the lower bound. As a result, a suitable bound should first be calculated before implementing the BNB algorithm.

\subsection{Simulated Annealing}\label{sec:SA}

SA is a stochastic optimisation technique for approximating the global optimum of a given objective function. This technique is based on the concept of annealing, which comes from metallurgy and statistical physics. Annealing is the idea of slowly cooling down a metal (initially at a high temperature) to ensure that the metal particles are orderly arranged to strengthen the metal \cite{burkard1998quadratic}. SA was designed to solve COP \cite{kirkpatrick1983optimization}. The algorithm commences with an initial random solution and a high temperature. A cooling schedule controls the decrease of the temperature \cite{kirkpatrick1983optimization}. SA is built on the Metropolis Algorithm \cite{hastings1970monte,metropolis1953equation} which is moderated by a decreasing temperature parameter. A slow decrease in the temperature allows SA to explore the search space irrespective of the quality of the solution. This decrease in the temperature helps to prevent local optimal solutions \cite{van1987simulated}. At each temperature, a neighbourhood search algorithm generates a solution. That updated solution's cost is compared to the rolling best solution. If the randomly generated solution is worse than the current best solution, the randomly generated solution becomes the new best solution. A Markov Chain is included in SA to moderate the temperature \cite{van1987simulated}. This process ends at ``the ground state'' (the lowest temperature value allowed), revealing the best solution found \cite{wilhelm1987solving}. 

\subsection{Variational Quantum Eigensolver}\label{sec:VQE}

The VQE algorithm is a quantum-classical hybrid algorithm that has two components, a quantum sub-routine and a classical loop \cite{peruzzo2014variational}. The VQE algorithm's objective is to find the ground state, or the minimum eigenvalue of the Hermitian matrix \cite{QiskitTextbook:2020}. The Hermitian matrices are suitable for describing the Hamiltonian $H$ of quantum systems, which represents the total energy of a considered system. The Hamiltonian system of a given COP can be described by:
\begin{equation}
    H = \Sigma_{\alpha}h_{\alpha}\otimes_{j = 1}^{N}\sigma_{\alpha_{j}}^{(j)},
\end{equation}
where $\sigma_{\alpha}$ is the Pauli matrix acting in the Hilbert space, $\otimes$ is a tensor product of Pauli matrices and $h_{\alpha}$ is the Hamiltonian coefficient.
The VQE is an iterative algorithm used to find the lowest eigenstate of the Hamiltonian system. The output of the quantum subroutine is used to update the ansatz. The ansatz of the VQE algorithm, which is classically prepared, are the parameterised trial states $\ket{\psi(\theta)}$. The quantum subroutine calculates and returns the expectations of the energy of the provided ansatz or the measured Hamiltonian. The expectation value is measured such that it satisfies the variational principle, which states that the expectation value is always greater than or equal to the ground state of the system \cite{peruzzo2014variational}. The Rayleigh-Ritz variational principle states: 
\begin{equation}
    \bra{\psi(\Vec{\theta)}}H\ket{\psi(\Vec{\theta)}} \leq \bra{\psi(\theta_0)}H\ket{\psi(\theta_0)} = E_0 \>,
    \label{eq:RRvar}
\end{equation}
where $\ket{\psi(\theta)}$ is any parametrised trial state and $E_0$ is the ground state. This principle ensures that VQE converges to the lowest possible eigenstate \cite{peruzzo2014variational}.

The variational method is a robust method for quantum computing, as unlike modern classical computing or theoretical quantum computing, NISQ technology is susceptible to significant errors as a result of noise \cite{QiskitTextbook:2020}. This noise is due to the sensitivity of qubits to the environment and hardware limitations of hardware \cite{murali2019noise}. However, even with accounting for noise, the VQE algorithm's solutions will not overshoot the minimum, provided the ansatz of $\Vec{\theta}$ includes the ground state \cite{moll2018quantum}. This is an advantage that makes the black-box optimiser, VQE, robust to noise.

Classical algorithms are applied to minimise the measured Hamiltonian state's expectation value by updating the parameter values. This occurs within the classical loop by finding the optimal set of $\theta$ parameters. Qiskit has an array of optimisers available, including COBYLA, LBRFGS and RBTOpt. The optimiser selected is based on the VQE circuit's depth, which is to be executed  \cite{QiskitTextbook:2020}. The VQE is a variational algorithm that favours short-depth circuits, which can be executed on NISQ devices \cite{Rattew2019ADN}. However, the VQE's heuristic nature does not guarantee the global optimal solution \cite{moll2018quantum}. The VQE algorithm’s performance depends on the variational form, a classical optimiser and the number of shots for each experiment. The variational form is the initial circuit that approximately determines the ground state of the Hamiltonian system. The type of circuits options available, on Qiskit, to represent the variational form are: \textsc{TwoLocal} (TL) circuit (linearly entangled, with an RY-gate layer) and the \textsc{RealAmplitudes} (RA) circuit (linear entanglement). A disadvantage of variational algorithms implemented on gate-model quantum computers like IBM's, is that they are predisposed to barren plateaus. Barren plateaus are where the gradient of the objective function exponentially decreases as a function of the number of qubits \cite{grant2019initialization}. These occur when using classical gradient descent methods, and there is no change in objective value \cite{mcclean2018barren}. A large circuit's average value of the gradient of the objective function is zero in barren plateaus, which means a solution cannot be obtained \cite{grant2019initialization}.


\subsection{Quantum Approximate Optimisation Algorithm}

The Quantum Approximate on Algorithm (QAOA) was introduced by \cite{farhi2014quantum} and, like the VQE, is based on the Variational Principle but has a unique trial state selection. Where VQE chooses a generic trial function $\ket{\psi(\theta)}$, the QAOA uses a Hamiltonian to build the trial function \cite{willsch2020benchmarking}. A Hamiltonian is used to described a cost function $C(x)$. The QAOA begins by finding a trial state $\ket{\psi_{p}(\gamma,\beta)}$ with real parameters $\gamma$ and $\beta$ such that the expectation value of the Hamiltonian is satisfied,
\begin{equation}
    \min F_{p}(\gamma,\beta) = \bra{\psi_{p}(\gamma,\beta)}H\ket{\psi_{p}(\alpha,\beta)} 
\end{equation} 
The trial state that depends on the parameters $\gamma$ and $\beta$, is prepared on a quantum computer and measured in the computational basis. Then a bit string $x^{n}$ is obtained. Continuous repetition of this process using the same value of $\gamma$ and $\beta$ will produce an optimal bit string $x^{*}$. This bit string, when measured in the computational basis, can be shown to have a high probability of minimising the cost function $C(x^{*})$. The integer, $p$, determines the circuit's depth level where each trial state is prepared. The depth level of the circuit is described by $p$ where $p \ge 1$ and the accuracy of the QAOA is said to improve as $p$ increases \cite{zhou2020quantum}.  Unlike VQE, QAOA has no initial variational form \cite{QiskitTextbook:2020}. Therefore, QAOA is less configured, making it potentially more effective at converging. On the other hand, QAOA using quantum computers to prepare parameters means it uses more quantum functions and exposes the results to more error \cite{QiskitTextbook:2020,streif2020training}.   

\setlength\extrarowheight{1.05pt}
\renewcommand{\arraystretch}{1.05}
\begin{table*}
\centering
\caption{Description of the IBM Basis Gates}\label{tab:quantum_gates}
    \begin{tabular}{|c|l|}
        \toprule\hline
        \textbf{IBM Devices' Gates} & \textbf{Description}\\ 
        \hline\hline
         Controlled NOT gate (CX) & Creates entanglement if target qubit is in superposition \\ 
         Identity gate (ID) & No operation is performed on a qubit  \\ 
         RZ & Rotation of qubit state around z axis\\
         Square Root of X gate (SX) & Creates superposition on qubit in downstate\\
         X & Changes state of qubit from down state to up state and vice versa \\
         \hline\bottomrule
    \end{tabular}
\end{table*}
\setlength\extrarowheight{1.0pt}
\renewcommand{\arraystretch}{1.00}

\subsection{New Set of Basis Gate on IBM Quantum Devices}

Quantum algorithms are often graphically represented as quantum circuits in the literature \cite{harrigan2021quantum}. Logic gates are used to describe the computational operations which change a state of a qubit in a quantum circuit \cite{QiskitTextbook:2020,zahedinejad2017combinatorial}. Some of the most common and basic quantum gates are the Hadamard gate and the Controlled NOT (CX) gate. The new set of basis gates on the IBM quantum devices currently consists of the (CX, ID, RZ, SX, X); these recently changed from  (CX, ID, U1, U2, U3) the original set of basis gates \cite{QiskitTextbook:2020}. The recent change to the IBM quantum gates is a conditional reset. The conditional reset feature initialises qubits in the $\ket{0}$ state through measuring, followed by a conditional NOT (CNOT) gate. Conditional reset allows faster circuit execution by reducing the initialisation time between shots - i.e. we would predict a decrease in computation time. Conditional reset also allows qubit reuse for circuits requiring a source of ``fresh ancillas". Ancillas moderate the interaction of qubits in a quantum register. Qubit reuse should make calculating circuits more efficient. IBM implemented state-discrimination logic in control systems to decode qubit states from received microwave pulses. Over 1 $\mu$s, the control identifies the qubit states and executes the qubit reuse. Enabling hardware state discrimination negates the necessity for discrimination calibration circuits (these were previously attached to each job). It, therefore, can be predicted that the conditional reset should improve the quality of results on the metrics investigated.

\section{Methodology}\label{sec:methodology}

\subsection{Experimental Settings}\label{sec:experimental}

An Intel$^{\tiny\text{\textregistered}}$ Core\texttrademark\ i5-6200U CPU @ 2.30GHz 4GB 64 bit Microsoft Windows 10 operating system was used to obtain the classic results on Python 3.6.5. Quantum algorithms are from the Aqua library on the open-source Python framework - \href{https://qiskit.org/}{Qiskit}. The IBM Quantum Experience Cloud was used to run all quantum experiments across various IBM simulators and quantum devices. The latest Qiskit version (0.23.3) allows access to devices with improved operational gates; however, to compare the performance of IBM quantum hardware on COP, an older version of Qiskit (0.20.0), in addition to the new version, was used. The programme for all experiments can be found on \href{https://github.com/QuCO-CSAM}{Github}.

\subsubsection{Algorithm Implementation}

The Qiskit framework has a library called Qiskit Aqua that supports quantum algorithms like the VQE and QAOA. The parameters associated with the VQE and QAOA have different functions that affect their overall performance when applied to a COP. The variational form is one such parameter. This parameter approximates a circuit that determines the ground state of a Hamiltonian system. The \textsc{TwoLocal} and \textsc{RealAmplitude} circuits were used in this research for the VQE. Unlike the VQE, where a variational form can be configured, the QAOA develops its own variational form. The Qiskit variational form for the QAOA depends on a $p$-value, which determines the parametrised circuit's depth. To increase optimality, a $p$-value of 3 is set for the QAOA on both the TSP and QAP. The number of shots, which provides a distribution of the results, is another parameter that affects the VQE and QAOA performance. The default number of shots for the VQE and QAOA is set at 1024 and 8192 respectively. Considering that the expectation value of a quantum state is optimised through classical optimisers, the Simultaneous Perturbation Stochastic Algorithm (SPSA) is recommended for variational quantum algorithms \cite{QiskitTextbook:2020}. Because of the various states that occur with probability, the state with the most probable feasible solution is chosen as the final solution. For the quantum experiments, a simulator, using the Matrix Product State (MPS), was used to find a good enough initial point that was used on the actual quantum devices. The MPS method uses a compact form to formulate two-qubit gate operations which implement the VQE \cite{QiskitTextbook:2020}. The strength of the initial point results in an estimation of solutions that are in the neighbourhood of the optimal solution and provides a higher probability of converging to the optimal solution, this is especially true for the VQE \cite{moll2018quantum}. Classical simulators do not experience as much noise compared to actual quantum devices hence obtaining an initial point through the simulators will increase the performance of the quantum algorithms when applied to actual quantum devices. These quantum devices (seen in Table~\ref{tab:quantum_devices}) are limited to the number of qubits they support (i.e. $n$ cities/locations uses $n^2$ qubits) as well as the depth of a circuit.

Using Discrete Optimisation CPLEX (DOcplex), the ``0-1" integer formulation of the QAP can be converted into a Quadratic Program. The Quadratic Program model can easily be mapped to a Hamiltonian using a Qiskit module. The ``0-1" integer formulation for the TSP is directly mapped to a Hamiltonian using a Qiskit module. The \textsc{BuildModel} function in Algorithms~\ref{al:VQE} and \ref{al:QAOA} represents these processes of building a Hamiltonian. The Variational Quantum Eigensolver (VQE) consists of three functions. The first is the fitness function using the ``0-1" formulation of the QAP. The following function is a feasibility verifier where the permutation matrix $x$ is checked against the constraints of the integer formulation of the QAP (a facility can only be assigned to one location and vice versa). Thirdly, the VQE has a function that returns all the feasible solution eigenstates and their accompanying probabilities. Since the VQE returns a distribution of probabilities for a suite of solutions, this function lists the returned eigenstates that are feasible along with their associated probability. This last function is called \textsc{FeasibleOutput} in Algorithm~\ref{al:VQE}. When using quantum computers as the backend, they are accessed through an IBM account, and the commands and results are sent via the internet. Algorithm~\ref{al:VQE} shows the implementation of VQE. 

Variational methods hold promise in solving optimisation problems given their potential to find the ground state of a Hamiltonian system, which encodes the optimisation problem. Therefore, the parameter selection for these methods is pertinent and should be further investigated in future work. For example, the impact of the employed $p$-value used in QAOA should be investigated. Furthermore, the emphasis placed on the initial solution used and its significance suggests that future work consider alternative techniques to find improved initial solutions.

 
\begin{algorithm}[H]
    \begin{algorithmic}[1]
        \Require
        \State{\texttt{inputmatrix(ces)}, \texttt{circ}, \texttt{initialpoint}, \texttt{maxiter}}
        \Ensure
        \State{\texttt{$\pi^*$}, \texttt{$cost^*$} }
        \State{initialisation;}
        \State{\texttt{qubitOpdocplex} = \textsc{BuildModel}(\texttt{inputmatrix(ces)})}
        \State{\texttt{num} = number of qubits of \texttt{qubitOpdocplex}} 
        \State{\texttt{spsa} = \textsc{SPSA}(\texttt{maxiter})}
        \If{\texttt{circ} = RA} 
            \State{\texttt{ry} = \textsc{RealAmplitudes}(\texttt{um}, \texttt{entanglement=linear})}
        \ElsIf{\texttt{circ} = TL}
            \State{\texttt{ry} = \textsc{TwoLocal}(\texttt{num}, \texttt{entanglement=linear)}}
        \EndIf
       
        \State{\texttt{vqe} = \textsc{VQE}(\texttt{qubitOpdocplex}, \texttt{ry}, \texttt{spsa}, \texttt{initialpoint})}
        \State{\texttt{quantuminstance} = \textsc{Backend}(1024 shots)}
        \State{\texttt{result} = \textsc{Run}(\texttt{vqe},\texttt{quantuminstance})} 
        \State{\texttt{$\pi^*$}, \texttt{$cost^*$} = \textsc{FeasibleOutput}(\texttt{result[eigenstate]})} 
    \end{algorithmic} 
    \caption{Variational Quantum Eigensolver (VQE)}
    \label{al:VQE}
\end{algorithm}

\begin{algorithm}[H]

    \begin{algorithmic}[1]
        \Require
        \State{\texttt{inputmatrix(ces)}, \texttt{initialpoint}, \texttt{maxiter}}
        \Ensure
        \State{\texttt{$\pi^*$}, \texttt{$cost^*$}}
        \State{initialisation;}
        \State{\texttt{qubitOpdocplex} = \textsc{BuildModel}(\texttt{inputmatrix(ces)})}
        \State{\texttt{num} = number of qubits of \texttt{qubitOpdocplex}}
        \State{\texttt{spsa} = \textsc{SPSA}(\texttt{maxiter})}
        \State{\texttt{qaoa} = \textsc{QAOA}(\texttt{qubitOpdocplex}, \texttt{spsa}, \texttt{initialpoint})}
        \State{\texttt{quantuminstance} = \textsc{Backend}(1024 shots)}
        \State{\texttt{result} = \textsc{Run}(\texttt{qaoa},\texttt{quantuminstance})}
        \State{\texttt{$\pi^*$}, \texttt{$cost^*$} = \textsc{FeasibleOutput}(\texttt{result[eigenstate]})}
    \end{algorithmic} 
    \caption{Quantum Alternating Optimisation Ansatz (QAOA)}
    \label{al:QAOA}
\end{algorithm}

\setlength \extrarowheight{1.5pt}
\renewcommand{\arraystretch}{1.05}
\begin{table*}[htbp]
	\centering
	\caption{Specifications of the IBM Quantum Devices\label{tab:quantum_devices}}
	\resizebox{0.95\textwidth}{!}{
		\begin{tabular}{|c|c|c|c|c|c|c|c|} 
			\toprule\hline
			\textbf{Quantum Device} & \textbf{Version} & \textbf{Available Qubits} & \textbf{Max Shots} & \textbf{Max Circuits} & \textbf{Processor Type} & \textbf{Average Readout Error} & \textbf{Average CNOT Error} \\  
			\hline\hline
			ibmq\_johannesburg       & 1.2.2         & 20                        & 8192               & 900   & Penguin R3 & 8.412e-2 &     -   \\ 
			\hline
			ibmq\_boeblingen         &  1.2.9        & 20                        & 8192               & 900   & Penguin R4 & 5.258e-2 &    -       \\
			\hline
			ibmq\_montreal           & 1.9.7          & 27                        & 8192               & 900      & Falcon R4 &  2.067e-2 &   1.010e-2            \\
			\hline
			ibmq\_cambridge          & 1.2.0         & 27                   & 8192               & 900 & Falcon R1.1 & 12.682e-2 &          -      \\
			\hline
			ibmq\_sydney             & 1.0.37         & 27                        & 8192               & 900     & Falcon R4 & 3.876e-2 & 1.291e-2            \\
			\hline
			ibmq\_toronto            & 1.4.22         & 27                        & 8192               & 900     & Falcon R4 & 4.253e-2&1.157e-2            \\
			\hline
			ibmq\_rochester          & 1.2.0         & 53                        & 8192               & 75       & Hummingbird R1  &  14.448e-2&     -     \\
			\hline
			ibmq\_manhattan          & 1.19.1          & 65                        & 8192               & 900            & Hummingbird R2 &3.843e-2&1.541e-2       \\
			\hline\bottomrule
		\end{tabular}
		}
    \begin{tablenotes}\footnotesize
    \item [-] Omitted Average CNOT Error values of decommissioned IBM quantum devices. Further details of these devices can be found \href{https://github.com/Qiskit/qiskit-terra/tree/master/qiskit/test/mock/backends}{here}.
    \end{tablenotes}
\end{table*}

\subsection{Metrics}

CPU time is used to measure the solution quality of the deterministic BNB algorithm used on the TSP and QAP. The stochastic nature of heuristic algorithms expands the number of performance measures that can test solution quality. In each experiment, 30 trials are executed, and two success rates (SR) and a feasibility percentage are recorded. The SR metric describes the percentage amount of the trials within 95\% and 99\% of the optimal global solution. The feasibility percentage details the percentage of trials that generate a feasible solution. This paper will also look into a metric that is specific to quantum results. As briefly mentioned in Section~\ref{sec:VQE}, the number of shots details the number of times a quantum algorithm is repeated. By default, the number of shots was set to 1024 for the VQE and QAOA. This implies that a quantum algorithm is repeated as many times as the shots for any particular instance. This leads to a distribution of all the results obtained for that specific instance. The optimal solution for a particular trial is the most probable feasible quantum eigenstate in that distribution. This paper expands that idea introduced in \cite{chieza2020computational} by considering a new metric called the \textit{uncertainty percentage}. This uncertainty percentage tests the quantum solution's quality by looking at how likely it is that solution can be obtained. An uncertainty percentage is calculated for each feasible solution in a set of 30 trials. This uncertainty percentage is based on the frequency of the quantum eigenstate. The uncertainty percentage analysis includes the average probability from the set, the maximum and minimum probability rates, and the standard deviation (note the headings in Table \ref{tab:newmetric}). In addition to the SR, feasibility rate and uncertainty percentage, the average CPU time is computed. IBM quantum devices are subject to calibration where a device is taken offline for performance testing. This does affect the CPU time on a device for any particular experiment. As a result, two types of CPU time performances are recorded: the average CPU time with outliers (AT) and the average CPU time without outliers (MT).


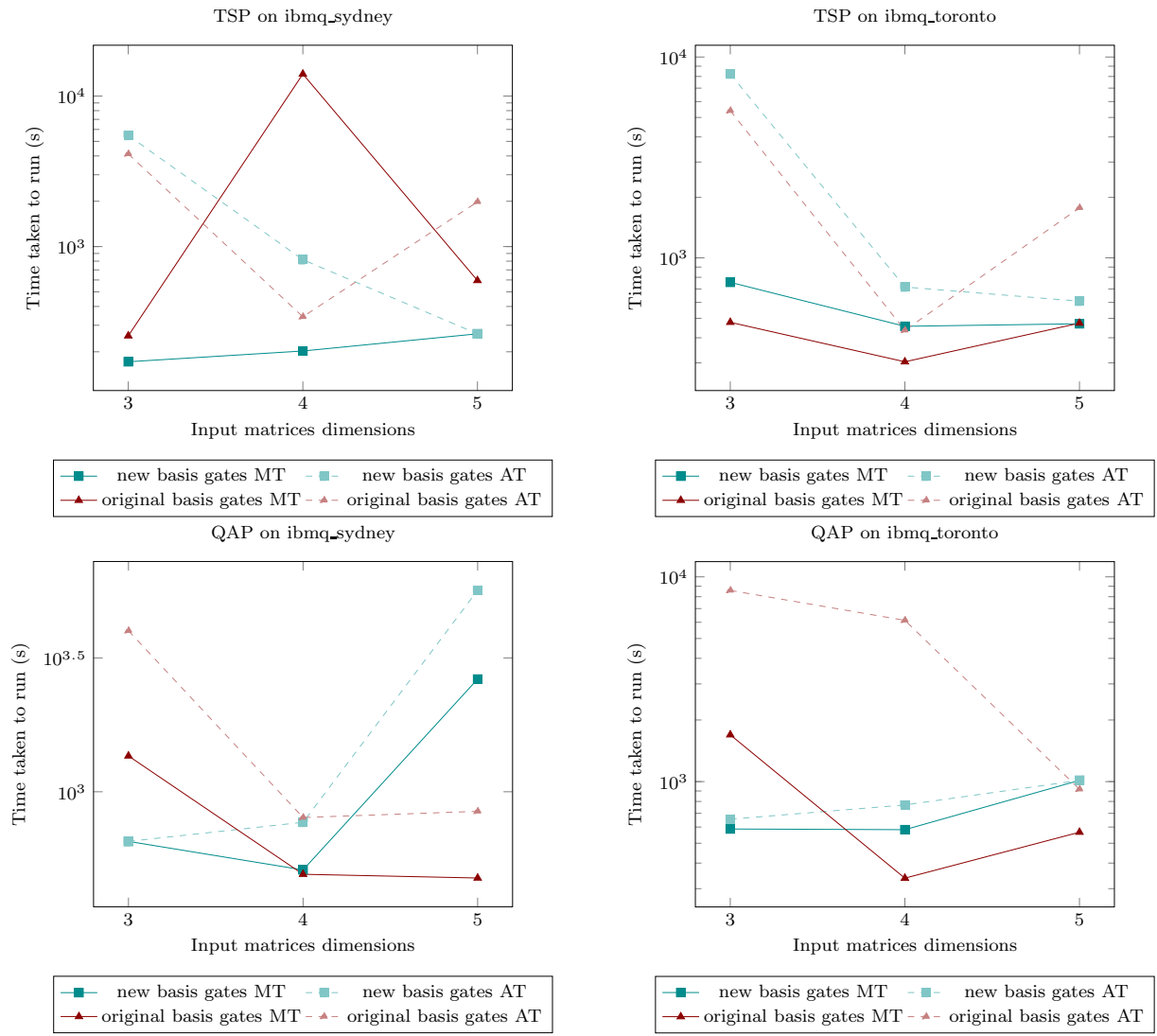
\begin{figure*}
    \centering
    \begin{subfigure}[t]{0.475\textwidth}
        \raggedleft
    \begin{tikzpicture}[scale=.85]
    	\begin{semilogyaxis}[
    	    xtick={3,4,5},
    		xlabel=Input matrices dimensions,
    		ylabel=Time taken to run (s),
    		title=TSP on ibmq\_sydney,
    		legend pos=north east,
    		legend style={at={(0.5,-0.2)},
		anchor=north,legend columns=2}
		]
    	    \addplot[color=darkcyan,mark = square*] coordinates {
    	    (3,171.65)
    	    (4,202.40)
    	    (5,263.23)
    	    };
    	    \addlegendentry{new basis gates MT}
    	    \addplot[color=darkcyan!50,mark = square*, dashed] coordinates {
    	    (3,5474.48)
    	    (4,817.91)
    	    (5,263.23)
    	    };
    	    \addlegendentry{new basis gates AT}
    	    \addplot[color=darkred,mark = triangle*, ] coordinates {
    	    (3,255.25)
    	    (4,14007.94)
    	    (5,594.93)
    	    };
    	    \addlegendentry{original basis gates MT}
    	    \addplot[color=darkred!50,mark = triangle*,dashed] coordinates {
    	    (3,4125.49)
    	    (4,341.72)
    	    (5,1982.69)
    	    };
    	    \addlegendentry{original basis gates AT}
    	\end{semilogyaxis}
    \end{tikzpicture}
    \label{fig:sydneyTSP}
    \end{subfigure}
    \begin{subfigure}[t]{0.475\textwidth}
        \raggedleft
    
    \begin{tikzpicture}[scale=.85]
    	\begin{semilogyaxis}[
    	    xtick={3,4,5},
    		xlabel=Input matrices dimensions,
    		ylabel=Time taken to run (s),
    		title=TSP on ibmq\_toronto,
    		legend pos=north east,
    		legend style={at={(0.5,-0.2)},
		anchor=north,legend columns=2}
		]
    	    \addplot[color=darkcyan,mark = square*] coordinates {
    	    (3,755.40)
    	    (4,456.22)
    	    (5,469.60)
    	    };
    	    \addlegendentry{new basis gates MT}
    	    \addplot[color=darkcyan!50,mark = square*, dashed] coordinates {
    	    (3,8226.18)
    	    (4,715.83)
    	    (5,608.42)
    	    };
    	    \addlegendentry{new basis gates AT}
    	    \addplot[color=darkred,mark = triangle*, ] coordinates {
    	    (3,477.47)
    	    (4,303.75)
    	    (5,473.06)
    	    };
    	    \addlegendentry{original basis gates MT}
    	    \addplot[color=darkred!50,mark = triangle*,dashed] coordinates {
    	    (3,5401.14)
    	    (4,436.00)
    	    (5,1774.34)
    	    };
    	    \addlegendentry{original basis gates AT}
    	\end{semilogyaxis}
    \end{tikzpicture}
    \label{fig:torontoTSP}
    \end{subfigure}
    \begin{subfigure}[t]{0.475\textwidth}
        \raggedleft
    \begin{tikzpicture}[scale=.85]
    	\begin{semilogyaxis}[
    	    xtick={3,4,5},
    		xlabel=Input matrices dimensions,
    		ylabel=Time taken to run (s),
    		title=QAP on ibmq\_sydney,
    		legend pos=north east,
    		legend style={at={(0.5,-0.2)},
		anchor=north,legend columns=2}
		]
    	    \addplot[color=darkcyan,mark = square*] coordinates {
    	    (3,652.57)
    	    (4,510.72)
    	    (5,2636.32)
    	    };
    	    \addlegendentry{new basis gates MT}
    	    \addplot[color=darkcyan!50,mark = square*, dashed] coordinates {
    	    (3,652.57)
    	    (4,768.95)
    	    (5,5659.09)
    	    };
    	    \addlegendentry{new basis gates AT}
    	    \addplot[color=darkred,mark = triangle*, ] coordinates {
    	    (3,1360.91)
    	    (4,491.77)
    	    (5,476.22)
    	    };
    	    \addlegendentry{original basis gates MT}
    	    \addplot[color=darkred!50,mark = triangle*,dashed] coordinates {
    	    (3,3994.68)
    	    (4,801.31)
    	    (5,845.03)
    	    };
    	    \addlegendentry{original basis gates AT}
    	\end{semilogyaxis}
    \end{tikzpicture}
    \label{fig:sydneyQAP}
    \end{subfigure}
    \begin{subfigure}[t]{0.475\textwidth}
        \raggedleft
    \begin{tikzpicture}[scale=.85]
    	\begin{semilogyaxis}[
    	    xtick={3,4,5},
    		xlabel=Input matrices dimensions,
    		ylabel=Time taken to run (s),
    		title=QAP on ibmq\_toronto,
    		legend pos=north east,
    		legend style={at={(0.5,-0.2)},
		anchor=north,legend columns=2}
		]
    	    \addplot[color=darkcyan,mark = square*] coordinates {
    	    (3,586.91)
    	    (4,582.44)
    	    (5,1014.09)
    	    };
    	    \addlegendentry{new basis gates MT}
    	    \addplot[color=darkcyan!50,mark = square*, dashed] coordinates {
    	    (3,655.30)
    	    (4,768.95)
    	    (5,1014.09)
    	    };
    	    \addlegendentry{new basis gates AT}
    	    \addplot[color=darkred,mark = triangle*, ] coordinates {
    	    (3,1692.50)
    	    (4,337.68)
    	    (5,566.01)
    	    };
    	    \addlegendentry{original basis gates MT}
    	    \addplot[color=darkred!50,mark = triangle*,dashed] coordinates {
    	    (3,8586.37)
    	    (4,6133.34)
    	    (5,917.79)
    	    };
    	    \addlegendentry{original basis gates AT}
    	\end{semilogyaxis}
    \end{tikzpicture}
    \label{fig:torontoQAP}
    \end{subfigure}
    
    \caption{These plots aim to illustrate the impact on CPU time that the new set of basis gates (CX, ID, RZ, SX, X) have on ibmq\_sydney and ibmq\_toronto for both methods solved with VQE for problems of sizes 3, 4 and 5. The problem instances used are symmetric and are inputs given as $n \times n$ matrices.  \label{fig:4gates}}
\end{figure*}

\begin{figure*}
\resizebox{0.95\textwidth}{!}{
\begin{tikzpicture}
\begin{semilogyaxis}[
    ybar,
    title=TSP VQE MT with the original basis gates,
    enlarge x limits=0.125,
    legend style={at={(0.5,0.82)},
      anchor=south, legend columns = 3},
    ylabel={Time (s)},
    xlabel={Problem Size ($n$)},
    symbolic x coords={3,4,5,6,7},
    xtick=data,
    nodes near coords align={vertical},
    bar width = 5.0 pt,
    log origin y=infty,
    grid=major,
    width=\textwidth,
    height=\axisdefaultheight,
    cycle list/Paired-11,
    cycle list shift=-5,
    grid style = {dashed, gray!30},
    legend style={at={(0.5,-0.25)},
		anchor=north,legend columns=7}
    ]
\addplot[draw=Paired-A,fill=Paired-A!70] coordinates  {(3,0.01)  (4,0.01)   (5,0.01)   (6, 0.01) (7, 0.03)};\addlegendentry{BNB};
\addplot[draw=Paired-B,fill=Paired-B!70] coordinates {(3,0.02)  (4,0.02)   (5,0.02)   (6, 0.02) (7,0.02)}; \addlegendentry{SA};
\addplot[draw=Paired-C,fill=Paired-C!70] coordinates{(3,49.34)   (4,103.42)  (5,214.86)   (6,0) (7,0)};\addlegendentry{ibmq\_qasm\_simulator};
\addplot[draw=Paired-D,fill=Paired-D!70] coordinates  {(3,1254.48) (4,181.85)  (5,0)  (6,0) (7,0)};\addlegendentry{ibmq\_johannesburg};
\addplot[draw=Paired-E,fill=Paired-E!70] coordinates {(3,421.42)  (4,358.26) (5,392.38)     (6,0) (7,0)};\addlegendentry{ibmq\_cambridge};
\addplot[draw=Paired-F,fill=Paired-F!70] coordinates {(3,1442.34) (4,414.31) (5,917.83)   (6,0) (7,0)};\addlegendentry{ibmq\_montreal};
\addlegendimage{empty legend};
\addlegendentry{};
\addplot[draw=Paired-G,fill=Paired-G!70] coordinates {(3,505.77)  (4,2084.45)   (5,1022.47)   (6, 2176.15) (7,0)};\addlegendentry{ibmq\_rochester};
\addplot[draw=Paired-H,fill=Paired-H!70] coordinates {(3,143.84)  (4,211.68)   (5,0)   (6,0) (7,0)};\addlegendentry{ibmq\_boeblingen};
\addplot[draw=Paired-I,fill=Paired-I!70] coordinates {(3,528.74)  (4,303.75)   (5,473.06)   (6,0) (7,0)};\addlegendentry{ibmq\_toronto};
\addplot[draw=Paired-J,fill=Paired-J!70] coordinates {(3,255.25)  (4,292.52)   (5,594.93)   (6,0) (7,0)};\addlegendentry{ibmq\_sydney};
\addplot[draw=Paired-L,fill=Paired-L!70] coordinates {(3,300.90)  (4,670.67)   (5,694.60)   (6, 691.01) (7,881.23)};\addlegendentry{ibmq\_manhattan};
\end{semilogyaxis}
\end{tikzpicture}
}
\caption{CPU times (s) for TSP (Log scale) \label{fig:TSPtime}}
\end{figure*}
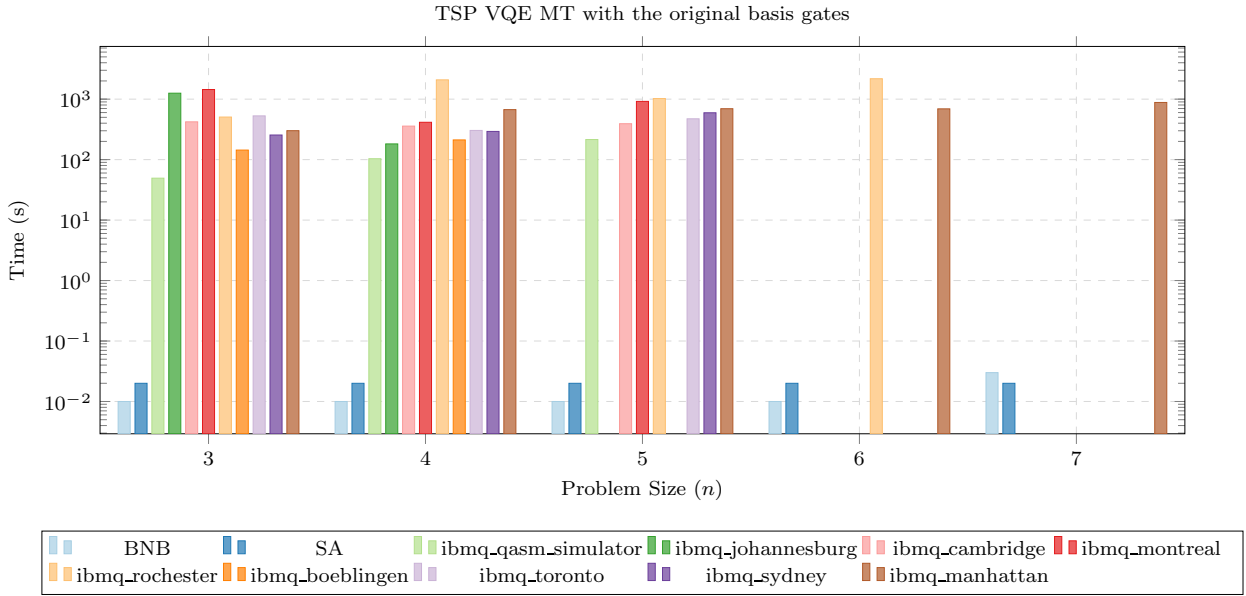
%

\begin{figure*}
\resizebox{.95\textwidth}{!}{
\begin{tikzpicture}
\begin{semilogyaxis}[
    ybar,
    title=QAP VQE MT with the original basis gates,
    enlarge x limits=0.125,
    legend style={at={(0.5,0.82)},
      anchor=south, legend columns = 3},
    ylabel={Time (s)},
    xlabel={Problem Size ($n$)},
    symbolic x coords={3,4,5,6,7},
    xtick=data,
    nodes near coords align={vertical},
    bar width = 5.0pt,
    log origin y=infty,
    width=\textwidth,
    height=\axisdefaultheight,
    cycle list/Paired-11,
    cycle list shift=-5,
    grid=major,
    grid style = {dashed, gray!30},
    legend style={at={(0.5,-0.25)},
		anchor=north,legend columns=7}
    ]
\addplot[fill=Paired-A!70,draw=Paired-A] coordinates {(3,0.01) (4,0.01) (5,0.01) (6,0.03) (7,0.28)};\addlegendentry{BNB};
\addplot[fill=Paired-B!70,draw=Paired-B] coordinates {(3,0.05) (4,0.11) (5,0.14) (6,0.16) (7,0.25)};\addlegendentry{SA};
\addplot[fill=Paired-C!70,draw=Paired-C] coordinates {(3,45.65) (4,45.64) (5,64.05) (6,) (7,)};\addlegendentry{ibmq\_qasm\_simulator};
\addplot[fill=Paired-D!70,draw=Paired-D] coordinates {(3,151.78) (4,342.20) (5,) (6,) (7,)};\addlegendentry{ibmq\_johannesburg};
\addplot[fill=Spectral-E!70,draw=Paired-E] coordinates {(3,400.48) (4,373.68) (5,872.01) (6,) (7,)};\addlegendentry{ibmq\_cambridge};
\addplot[fill=Paired-F!70,draw=Paired-F] coordinates {(3,283.07) (4,367.47) (5,565.01) (6,) (7,)};\addlegendentry{ibmq\_montreal};
\addlegendimage{empty legend}
\addlegendentry{}
\addplot[fill=Paired-G!70,draw=Paired-G] coordinates {(3,1702.85) (4,1133.78) (5,958.51) (6,2796.73) (7,)};\addlegendentry{ibmq\_rochester};
\addplot[fill=Paired-H!70,draw=Paired-H] coordinates {(3,268.71)  (4,873.04)   (5,)   (6,) (7,)};\addlegendentry{ibmq\_boeblingen};
\addplot[fill=Paired-I!70,draw=Paired-I] coordinates {(3,1692.50)  (4,337.68)   (5,566.01)   (6,) (7,)};\addlegendentry{ibmq\_toronto};
\addplot[fill=Paired-J!70,draw=Paired-J] coordinates {(3,1360.91)  (4,491.77)   (5,476.22)   (6,) (7,)};\addlegendentry{ibmq\_sydney};
\addplot[fill=Paired-L!70,draw=Paired-L] coordinates {(3,557.68)  (4,267.96)   (5,1318.44)   (6, 2967.89) (7,7928.29)};\addlegendentry{ibmq\_manhattan};


\end{semilogyaxis}
\end{tikzpicture}
}
\caption{CPU times (s) for QAP (Log scale)\label{fig:QAPtime}}
\end{figure*}
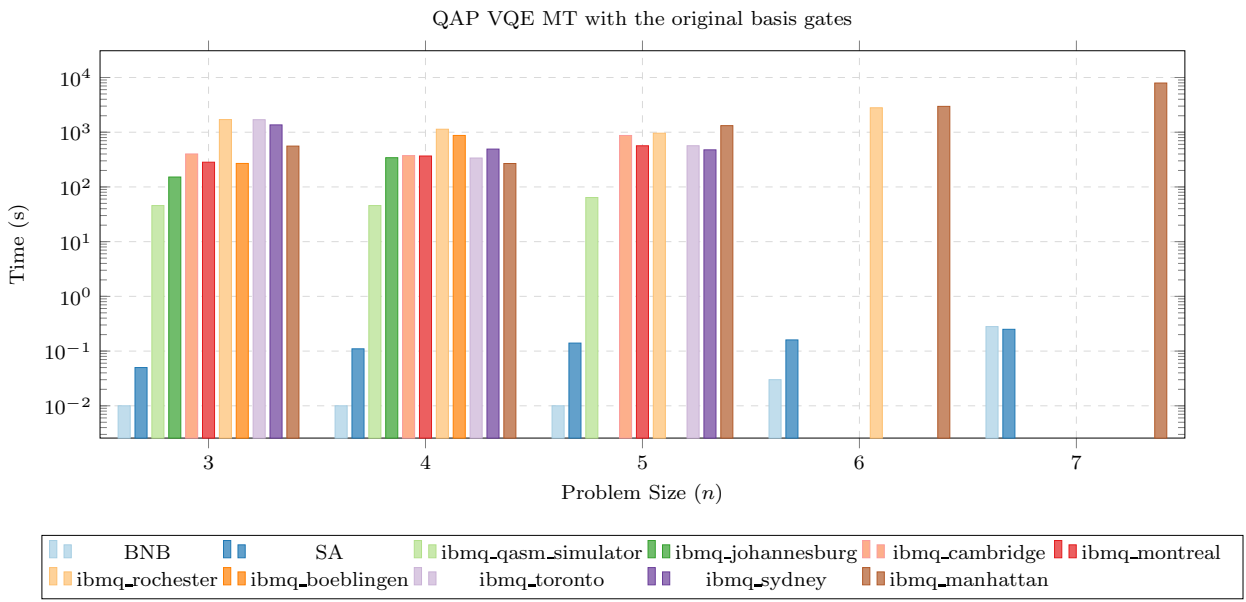

\section{Results}\label{sec:results}


The performance of the quantum optimisation algorithms benchmarked against classical optimisation algorithms, presented in Table~\ref{tab:results}, \ref{tab:results2} and \ref{tab:newmetric}, and Figures~\ref{fig:TSPtime}, \ref{fig:QAPtime} and \ref{fig:4gates}, are analysed with respect to both solution quality and computational time performance in this section.    

The experimental results show that classical optimisation algorithms outperform quantum optimisation algorithms with respect to success rate, feasibility and computational time in finding solutions to the TSP and QAP.


The solution quality metrics results reported for problem instances of sizes 3 and 4 on the TSP and QAP suggest that the VQE performs better when compared with the QAOA. The solution quality metric results obtained for conditionally reset devices do not have a noticeable difference compared to the metrics of IBM devices that were not conditionally reset.

IBM has introduced the conditional reset feature in order to make the quantum device calculations more efficient. The results in Table~\ref{tab:results2} use the gates explained in Table~\ref{tab:quantum_gates} on the devices available with conditional reset on VQE and QAOA. Figure~\ref{fig:4gates} compares the impact of conditional reset on time and for both average times AT and MT on both devices. From Figure~\ref{fig:4gates}, no clear improvement in computational time can be observed between results with or without conditional reset. Furthermore, when comparing the success rates in Table~\ref{tab:results2} to those on the same devices without conditional reset in Table~\ref{tab:results}, there is no improvement in success rate or feasibility. We, therefore, conclude that the implementation of conditional reset did not impact the results obtained.


Comparing the computational time taken to obtain a solution using classical algorithms compared to the VQE algorithm, it was found that the applied classical algorithms had a faster computational time performance than the VQE algorithm. Figures~\ref{fig:TSPtime} and \ref{fig:QAPtime} illustrate that the performance of the various employed quantum devices is consistent in terms of computational time, with the simulator performing the best. There is no distinct correlation between problem size and computational time for the quantum devices. However, this claim is made with a limited number of problem instances.


Success rate and feasibility are higher in SA than in VQE for most instances of the TSP and QAP, indicating that the SA obtains better solution quality than the VQE. From Table~\ref{tab:results}, it is evident that the success rate and feasibility percentage decrease as the problem size increase. Instances of sizes 3 for the TSP applied to the VQE perform the best in terms of feasibility and success rate out of all the TSP instances. This is likely because the TSP has the least number of operations (Table~\ref{tab:formulations}). Instances of sizes 6 and 7 perform the worst with the most operations and largest parameters. ibmq\_qasm-simulator has a higher feasibility percentage than the quantum devices. Comparing success rate and feasibility between devices shows no device has a superior performance to the others.

Comparing the results of the QAOA to those of VQE in Table~\ref{tab:results2}, the QAOA takes a longer computational time performance on the quantum devices than the VQE. This increased computational time is likely due to increased operations and depth (Table~\ref{tab:formulations}). Nevertheless, the success rate and feasibility of the QAOA are comparable to that of VQE. However, it is difficult to make inferences on the QAOA due to the size limitations of the instances and the lack of devices.



Table~\ref{tab:newmetric} aims to describe the quality of the feasible trials obtained using the quantum heuristics. Quantum circuits do not return a single answer but a series of eigenstates with corresponding frequencies. The corresponding percentage describes the likelihood of the most probable feasible solution is to be in the result of a given trial. This is unique to the success rate, as it measures the uncertainty of the current trial and not the probability of success for the heuristic. This uncertainty percentage is low across all instances, indicating that most of the returned eigenstates are infeasible and highlight the detriment of noise to solution quality. This accentuates the gap in performance between simulators and the ideal solutions from theoretical quantum computers and the applied algorithms' performance on NISQ devices. There is a correlation between the size of an instance and the uncertainty percentage. Results for VQE and QAOA are comparable when comparing instances of the same size. The results obtained using the VQE algorithm with and without conditional reset are comparable when comparing instances of the same size. Conclusions about the uncertainty percentage cannot be drawn for larger problem instances, as the set of feasible solutions is empty or singular.
The NISQ devices are limited by their specifications leading to a limit in the size of the problems that can be solved. Table~\ref{tab:quantum_devices} shows the limited access to qubits that prevents the larger instances from being tested. The other specifications impact error in the results and the computational time is taken to solve circuits. Table \ref{tab:formulations} shows how the size of parameters, number of operations and circuit depth grow with problem size. This phenomenon limits experiments to small instances. Therefore, for the QAOA on both the TSP and QAP, the experiments were run until size instance 4. Instances greater than size 4 cannot be simulated in polynomial time because of the circuit depth and the number of operations produced by those instances. Obtaining a good enough initial point for instances greater than size 4 is intractable on NISQ devices. QAOA uses more operations than VQE for problems of the same size, which may impact why it performs worse than VQE on all metrics. 



\begin{table*}[htbp]
\centering
\caption{Specifications of the circuits before transpiling on the quantum devices \label{tab:formulations}}
\begin{threeparttable}
\begin{tabular}{|c|c|c|c|c|c|c|c|c|}
\toprule\hline
\textbf{Size} & \textbf{Qubits} & \multicolumn{3}{|c|}{\textbf{TSP}} & \multicolumn{3}{|c|}{\textbf{QAP}}  & \textbf{Method}\\
\hline
 & & \textbf{Par}.\tnote{$\dagger$} & \textbf{Operations} & \textbf{Depth} &  \textbf{Par}.\tnote{$\dagger$} & \textbf{Operations} & \textbf{Depth} &\\
\hline\hline
\multirow{2}{*}{3}  & \multirow{2}{*}{9} & [100, ``TL"]  & 103 & 23 & [1000, ``TL"] & 103 & 23 & VQE\\
\cline{3-9}
 &  & [50] & 396 &116  & [100] & 396 & 149  & QAOA\\
\hline
 \multirow{2}{*}{4} & \multirow{2}{*}{16} & [1100, ``TL"] & 187 & 30 & [3000, ``TL"] & 187 & 30 & VQE\\
\cline{3-9}
 &  & [100] &992  &182  & [50] & 992 & 218  & QAOA\\
\hline
\multirow{2}{*}{5}  & \multirow{2}{*}{25} & [5500, ``RA"]  & 197 &  33& [5000, ``TL"] & 295 & 39 & VQE\\
\cline{3-9}
 &  & - &  2000& 260 & - & 1820 & 227  & QAOA\\
\hline
\multirow{2}{*}{6}  & \multirow{2}{*}{36} & [5000, ``TL"]  & 427 & 50 & [6000, ``TL"] & 427 & 50 & VQE\\
\cline{3-9}
 &  & - & 3528 & 359 & - & 3528 & 312  & QAOA\\
\hline
\multirow{2}{*}{7}  & \multirow{2}{*}{49} & [10000, ``TL"]  & 583 & 63 & [12000, ``TL"] & 583 & 63 & VQE\\
\cline{3-9}
 &  & - & 5684   &  565 & - & 6818 & 540 & QAOA\\
\hline
\multirow{2}{*}{8}  & \multirow{2}{*}{64} & - & 763 &  78& - & 763 & 78 & VQE\\
\cline{3-9}
 &  & - &  8576 & 590 & - & 10088 & 728  & QAOA\\
\hline
\multirow{2}{*}{9}  & \multirow{2}{*}{81} & - & 967 & 95 & - & 967 & 95 & VQE\\
\cline{3-9}
 &  & - & 12312 & 722 & - & 15552 & 970 & QAOA\\
\hline\bottomrule
\end{tabular}
\begin{tablenotes}\footnotesize
\item[$\dagger$] For VQE - [number of SPSA trials, variational form],
\item [] For QAOA - [number of SPSA trials]
\end{tablenotes}
\end{threeparttable}
\end{table*}


\setlength\extrarowheight{1.5pt}
\begin{table*}[htbp]
\caption{Experimental Results: quantum results with the original basis gates and all classical results \label{tab:results}}
\begin{threeparttable}
\begin{center}
\resizebox{.975\textwidth}{!}{
\begin{tabular}{|c|c|c|c|c|c|c|c|c|c|c|c|c|c|c|c|}
\toprule\hline
\textbf{Size} & \multicolumn{6}{|c|}{\textbf{TSP}} & \multicolumn{6}{|c|}{\textbf{QAP}} & \textbf{Devices} & \textbf{Method} \\
\hline
 & \textbf{Par}.\tnote{$\dagger$} & \textbf{SR99} & \textbf{SR95} & \textbf{Feas}.\tnote{*} & \textbf{AT(s)}\tnote{*} & \textbf{MT(s)}\tnote{*} & \textbf{Par}.\tnote{$\dagger$} & \textbf{SR99} & \textbf{SR95} & \textbf{Feas}.\tnote{*} & \textbf{AT(s)}\tnote{*} & \textbf{MT(s)}\tnote{*}& &\\
\hline\hline
\multirow{11}{*}{3} & \multirow{9}{*}{[100, ``TL"]}&100.0  &100.0 & 100.0 &43.33&43.33  & \multirow{9}{*}{[1000, ``TL"]}&100.0 & 100.0 & 100.0& 45.65 & 45.65 & ibmq\_qasm-simulator & \multirow{9}{*}{VQE} \\
\cline{3-7} \cline{9-14}
 & &100.0 &100.0 & 100.0 & 2909.70&1771.32 & & 86.67 & 86.67 & 100.0 & 1486.68 & 151.78 &ibmq\_johannesburg & \\
\cline{3-7} \cline{9-14}
 & & 100.0 & 100.0& 100.0&1981.35&1485.74 & & 43.33 & 43.33 & 100.0 & 479.18 & 415.05 & ibmq\_montreal & \\
\cline{3-7} \cline{9-14}
 & &100.0 &100.0& 100.0& 423.84& 423.84& & 96.67 & 96.67 & 100.0 & 414.38 & 520.24 & ibmq\_cambridge &\\
\cline{3-7} \cline{9-14}
 & &100.0 & 100.0&100.0&7271.66 &1494.03 & & 53.33 & 53.33 & 100.0 & 987.22 & 548.20 & ibmq\_rochester& \\
\cline{3-7} \cline{9-14}
 & & 100.0 &100.0 & 100.0& 201.78 & 143.84 & & 0.0 & 0.0 & 100.0 & 852.69 & 268.71 & ibmq\_boeblingen &  \\
\cline{3-7} \cline{9-14}
 & & 100.0 &100.0 & 100.0&4125.49 & 255.25 & & 93.33 & 93.33 & 100.0 & 3994.88 & 1360.91 & ibmq\_sydney &  \\
\cline{3-7} \cline{9-14}
 & &100.0  &100.0 &100.0 & 5401.14&  528.74& & 73.33 & 73.33 & 100.0 & 8586.37 & 1692.50 & ibmq\_toronto &  \\
\cline{3-7} \cline{9-14}
 & &100.0  &100.0 &100.0 &530.38 &300.90  & & 96.67 & 96.67 & 100.0 & 3722.58 & 557.68 & ibmq\_manhattan &  \\
\cline{2-15}
 &[0.01, 10, 0.8, 10] &100.0 &100.0& 100.0&0.02 &0.02 & [1.0, 20, 0.90, 20] & 100.0 & 100.0& 100.0 & 0.05 & 0.05& classical-device& SA \\
\cline{2-15}
 & -& -&-&- &0.01 & 0.01& - &  -&  -&- & 0.01 & 0.01 &  classical-device& BNB \\
\hline
\multirow{11}{*}{4} & \multirow{9}{*}{[1100, ``TL"]} &3.33 &3.33 & 73.33& 54.46&54.46 &\multirow{9}{*}{[3000, ``TL"]} & 0.0 & 100.0 & 100.0 & 45.64 & 45.64 &ibmq\_qasm-simulator & \multirow{9}{*}{VQE}\\
\cline{3-7} \cline{9-14}
 & & 23.33& 23.33&56.67 & 215.49& 215.49& & 3.33 & 16.67 & 56.67 & 495.10 & 342.20 &ibmq\_johannesburg &\\
\cline{3-7} \cline{9-14}
 & &13.33 &13.33&53.33 & 681.85& 445.32& & 3.33 & 6.67 & 56.67 & 827.51 & 369.81 &ibmq\_montreal &\\
\cline{3-7} \cline{9-14}
 & & 16.67&16.67&43.33 & 365.35& 365.35& & 6.67 & 13.33 & 63.33 & 562.17 & 367.47 &ibmq\_cambridge &\\
\cline{3-7} \cline{9-14}
& & 23.33 & 0.0&43.33&2006.08& 1276.74& & 6.67 & 13.33 & 43.33 & 4803.45 & 3260.97 &ibmq\_rochester &\\
\cline{3-7} \cline{9-14}
 & & 6.67 & 6.67&30.0 &230.62 & 211.68 & & 0.0 & 10.0 & 46.67 & 1397.01 & 873.04 & ibmq\_boeblingen &  \\
\cline{3-7} \cline{9-14}
 & & 3.33 &3.33 &6.67 &341.72 & 292.52 & & 0.0 & 3.33 & 36.67 & 801.31 & 491.77 & ibmq\_sydney &  \\
\cline{3-7} \cline{9-14}
 & &16.67  & 16.67& 40.0&436.00 & 303.75 & & 0.0 & 6.67 & 40.0 & 6133.34 & 337.68 & ibmq\_toronto &  \\
\cline{3-7} \cline{9-14}
 & &23.33  &0.0 &33.33 &2487.12 & 670.67 & & 0.0 & 10.0 & 40.0 & 298.94 & 267.96 & ibmq\_manhattan &  \\
\cline{2-15}
 &[0.01, 10, 0.8, 10] & 100.0&100.0&100.0 &0.02& 0.02& [1.0, 20, 0.90, 20] & 100.0 & 100.0& 100.0 & 0.11 & 0.11 & classical-device& SA \\
\cline{2-15}
 & -& -&-& &0.01& 0.01& -& -& -&& 0.01 & 0.01 & classical-device& BNB \\
\hline
\multirow{9}{*}{5} & \multirow{7}{*}{[5500, ``RA"]} &26.67 &26.67&26.67 & 99.51&92.17 & \multirow{7}{*}{[5000, ``TL"]} & 0.0 & 0.0 & 96.67 & 64.05 & 64.05 &ibmq\_qasm-simulator & \multirow{7}{*}{VQE}\\
\cline{3-7} \cline{9-14}
 & & 0.0 & 0.0& 0.0&562.13&388.65 & & 0.0 & 0.0 & 0.0 & 9986.32 & 872.01 &ibmq\_montreal& \\
\cline{3-7} \cline{9-14}
 & &0.0 &0.0&0.0 &1135.77& 439.38 & & 0.0 & 0.0 & 0.0 & 565.01 & 565.01 &ibmq\_cambridge& \\
\cline{3-7} \cline{9-14}
 & &3.33&3.33& 3.33& 13137.19 & 1228.91& & 0.0 & 0.0 & 0.0 & 2116.71 & 1801.51 &ibmq\_rochester &\\
 \cline{3-7} \cline{9-14}
 & &0.0  &0.0 &3.33 & 1982.69& 594.93 & & 0.0 & 0.0 & 0.0 & 845.03 & 476.22 & ibmq\_sydney &  \\
\cline{3-7} \cline{9-14}
 & & 0.0 &0.0 & 3.33&1774.34 & 473.06 & & 3.33 & 3.33 & 6.67 & 917.79 & 566.01 & ibmq\_toronto &  \\
\cline{3-7} \cline{9-14}
 & & 0.0 &0.0 &3.33 & 1836.56&694.60  & & 0.0 & 0.0 & 3.33 & 3020.28 & 1318.44 & ibmq\_manhattan &  \\
\cline{2-15}
 &[0.01, 10, 0.8, 10] &100.0 &100.0& 100.0 &0.02& 0.02& [1.0, 20, 0.90, 20] & 100.0 & 100.0& 100.0 & 0.14 & 0.14 & classical-device&SA \\
\cline{2-15}
 & -&- &-&- &0.01&0.01 & -& -& -&-& 0.01 & 0.01 & classical-device&BNB \\
\hline
\multirow{4}{*}{6} & \multirow{2}{*}{[5000, ``TL"]} &0.0 &0.0&0.0 &2476.87& 2176.15&\multirow{2}{*}{[6000, ``TL"]} & 0.0 & 0.0 & 3.33 & 5017.53 & 2265.98 & ibmq\_rochester & \multirow{2}{*}{VQE}\\
\cline{3-7} \cline{9-14}
 & & 0.0 & 0.0&0.0 &752.17 & 691.01 & & 0.0 & 0.0 & 0.0 & 24373.16 & 2967.89  & ibmq\_manhattan &  \\
 \cline{2-15}
 &[0.01, 10, 0.8, 10] & 100.0& 100.0&100.0 &0.02&0.02 & [1.0, 20, 0.90, 20] & 60.0 & 100.0& 100.0 & 0.16 & 0.16 & classical-device& SA \\
\cline{2-15}
 & -& -&-&- & 0.01& 0.01& -& -& -&-&0.03 & 0.03 & classical-device&BNB \\
 \hline 
\multirow{3}{*}{7} & [10000, ``TL"] &0.0 &0.0&0.0&2525.73 &881.23 & [12000, ``TL''] & 0.0 & 0.0 & 0.0 & 18429.26 & 7928.29  &ibmq\_manhattan & VQE \\
 \cline{2-15}
& [0.01, 10, 0.8, 10]& 66.67&66.67& 100.0&0.02&0.02 & [1.0, 20, 0.90, 740] & 93.3 & 93.3& 100.0 & 0.25 & 0.25 & classical-device&SA \\
\cline{2-15}
 &- &- &-&- &0.03& 0.03& -& -& -&-& 0.28 & 0.28 &classical-device &BNB \\
\hline\bottomrule
\end{tabular}
}
\end{center}
\begin{tablenotes}\footnotesize
\item[*] Feas. - percentage feasible: AT - average time with outliers, MT - average time without outliers
\item[$\dagger$] For VQE - [number of SPSA trials, variational form],
\item [] For SA - [tolerance, Markov chain length, cooldown factor, starting temperature] 
\end{tablenotes}
\end{threeparttable}
\end{table*}


\setlength\extrarowheight{1.5pt}
\begin{table*}[htbp]
\caption{Experimental quantum results with new basis gates \label{tab:results2}}
\begin{threeparttable}
\begin{center}
\resizebox{\textwidth}{!}{
\begin{tabular}{|c|c|c|c|c|c|c|c|c|c|c|c|c|c|c|c|}
\toprule\hline
\textbf{Size} & \multicolumn{6}{|c|}{\textbf{TSP}} & \multicolumn{6}{|c|}{\textbf{QAP}} & \textbf{Devices} & Method\\
\hline
 & \textbf{Par}.\tnote{$\dagger$} & \textbf{SR99} & \textbf{SR95} & \textbf{Feas}.\tnote{*} & \textbf{AT(s)}\tnote{*} & \textbf{MT(s)}\tnote{*} & \textbf{Par}.\tnote{$\dagger$} & \textbf{SR99} & \textbf{SR95} & \textbf{Feas}.\tnote{*} & \textbf{AT(s)}\tnote{*} & \textbf{MT(s)}\tnote{*}& &\\
\hline\hline
\multirow{5}{*}{3}  & \multirow{2}{*}{[100, ``TL"]} &100.0  & 100.0 & 100.0 &5474.48 & 171.65 & \multirow{2}{*}{[1000, ``TL"]} & 86.67 & 86.67 & 100.0 & 652.57 & 652.57 & ibmq\_sydney & \multirow{2}{*}{VQE} \\
\cline{3-7} \cline{9-14}
 & & 100.0 &100.0  &100.0  & 8226.18  & 477.47 & &  30.0 & 30.0 & 100.0 & 655.30 & 586.91  & ibmq\_toronto &\\
\cline{2-15}
 & \multirow{3}{*}{[50]}& 90.00 &90.0 &90.0 &124.59&124.59  & \multirow{3}{*}{[100]}& 30.0 & 30.0 & 100.0 & 62.57 & 62.57 & ibmq\_qasm-simulator & \multirow{3}{*}{QAOA} \\
\cline{3-7} \cline{9-14}
 & &100.0  & 100.0&100.0 & 3856.01&  730.30& -& -& - &- &  -& - & ibmq\_sydney &  \\
\cline{3-7} \cline{9-14}
 & &100.0  & 100.0& 100.0&10871.32 &755.40  & & 0.0 & 0.0 & 100.0 & 552.26 & 453.37 & ibmq\_toronto &  \\
\hline
 \multirow{5}{*}{4} & \multirow{2}{*}{[1100, ``TL"]} & 10.0 & 10.0& 36.67& 817.91&202.40  & \multirow{2}{*}{[3000, ``TL"]} & 3.33 & 6.67 & 36.67 & 768.95 & 510.72 & ibmq\_sydney  & \multirow{2}{*}{VQE} \\
\cline{3-7} \cline{9-14}
 & &6.67  &6.67 &33.33 &715.83 & 456.22 & & 0.0 & 3.33 & 30.0 & 720.66 & 582.44 & ibmq\_toronto  & \\
 \cline{2-15}
 & \multirow{3}{*}{[100]}& 0.0 & 0.0&0.0 & 164.05& 164.05 & \multirow{3}{*}{[50]}& 0.0 & 0.0 & 23.33 & 140.96 & 140.96 & ibmq\_qasm-simulator & \multirow{3}{*}{QAOA} \\
\cline{3-7} \cline{9-14}
 & &0.0  &20.0&43.33  &14007.94 & 901.45 & -& -&  -& -&  -&  -& ibmq\_sydney &  \\
\cline{3-7} \cline{9-14}
 & & 23.33 &23.33 &43.33 &2325.83 & 1055.58 & & 0.0 & 6.67 & 36.67 & 6979.90 & 1446.55 & ibmq\_toronto &  \\
\hline
\multirow{2}{*}{5} & \multirow{2}{*}{[5500, ``RA"]} &0.0  &0.0 &3.33 & 263.23& 263.23 & \multirow{2}{*}{[5000, ``TL"]} & 0.0 & 0.0 & 0.0 & 5659.09 & 2636.32 & ibmq\_sydney & \multirow{2}{*}{VQE}  \\
\cline{3-7} \cline{9-14}
 & & 0.0 & 0.0& 0.0& 608.42& 469.60 & & 0.0 & 0.0 & 3.33 & 1014.09 & 1014.09 & ibmq\_toronto &  \\
\hline\bottomrule
\end{tabular}
}
\end{center}
\begin{tablenotes}\footnotesize
\item[*] Feas. - percentage feasible: AT - average time with outliers, MT - average time without outliers
\item[$\dagger$] For VQE - [number of SPSA trials, variational form],
\item [] For QAOA - [number of SPSA trials]
\end{tablenotes}
\end{threeparttable}
\end{table*}

%

\setlength\extrarowheight{1.5pt}
\begin{table*}[htbp]
\caption{Experimental Results: describing the uncertainty percentage metric \label{tab:newmetric}}
\begin{threeparttable}
\begin{center}
\resizebox{.975\textwidth}{!}{
\begin{tabular}{|c|c|c|c|c|c|c|c|c|c|c|c|c|c|c|}
\toprule\hline
\textbf{Size} & \multicolumn{5}{|c|}{\textbf{TSP}} & \multicolumn{5}{|c|}{\textbf{QAP}} & \textbf{Devices}& \textbf{CR}.\tnote{$\dagger$} & \textbf{Method} \\
\hline
 & \textbf{\# feas.}\tnote{$\dagger$} & \textbf{mean (\%)} & \textbf{max (\%)} & \textbf{min (\%)} & \textbf{std (\%)} & \textbf{\# feas.}\tnote{$\dagger$} & \textbf{mean (\%)} & \textbf{max (\%)}& \textbf{min (\%)} & \textbf{std (\%)}& & &\\
\hline\hline
\multirow{14}{*}{3} &30 & 22.36& 99.41& 0.10 &35.35 & 30 & 95.65 & 98.93 & 0.68 & 17.64 &  ibmq\_qasm-simulator& F & \multirow{11}{*}{VQE} \\
\cline{2-13}
 & 30& 0.70& 1.75& 0.29 &0.35 & 30 & 0.57 & 1.07 & 0.20 & 0.22 &ibmq\_johannesburg& F &\\
\cline{2-13}
 &30 & 0.52& 0.97& 0.19 &0.14 & 30 & 0.41 & 0.78 & 0.20 & 0.13 & ibmq\_montreal &F& \\
\cline{2-13}
 & 30& 1.14&3.03 &0.20  & 0.96& 30 & 2.79 & 4.69 & 0.59 & 1.22 & ibmq\_cambridge &F& \\
 \cline{2-13}
 & 30&0.01 & 0.01& 0.01 &0.0 & 30 & 0.55 & 1.07 & 0.20 & 0.21 & ibmq\_rochester&F& \\
\cline{2-13}
 & 30&0.41 &0.59 & 0.20 &0.12 & 30 & 0.64 & 0.98 & 0.39 & 0.18 & ibmq\_boeblingen &F &  \\
\cline{2-13}
 &30 & 0.52& 0.98& 0.20 & 0.20& 30 & 1.04 & 2.54 & 0.20 & 0.48 &   ibmq\_sydney &F & \\
\cline{2-13}
 & 30& 0.52&0.88 & 0.20 &0.17 & 30 & 0.59 & 0.98 & 0.20 & 0.21 &  ibmq\_toronto &F & \\
\cline{2-13}
 &30 &6.02 & 24.51&  0.20& 7.27& 30 & 28.62 & 35.25 & 0.29 & 5.96 &  ibmq\_manhattan & F&  \\
 \cline{2-13}
 &30 &0.53 &1.17 &0.20  &0.22 & 30 & 0.77 & 2.05 & 0.20 & 0.43 &  ibmq\_sydney &T&  \\
\cline{2-13}
 &30 & 0.48& 0.78& 0.20 &0.13 & 30 & 0.49 & 0.88 & 0.20 & 0.16 & ibmq\_toronto &T&  \\
\cline{2-14}
 & 27&0.33 &0.88 &0.10  &  0.21
& 30 & 0.47 & 1.07 & 0.10 & 0.25 & ibmq\_qasm-simulator& T  & \multirow{3}{*}{QAOA} \\
 \cline{2-13}
 &30 & 0.36& 0.68& 0.20 &0.11 & - & -& -&  -&  -& ibmq\_sydney & T& \\
\cline{2-13}
 &30 & 0.41&0.68 &0.20  &0.13 & 30 & 0.39 & 0.68 & 0.2 & 0.13 &  ibmq\_toronto & T& \\
\hline
\multirow{14}{*}{4} &22 &32.51 &98.63 &0.10 & 30.83& 30 & 96.68 & 98.34 & 48.63 & 8.92 & ibmq\_qasm-simulator & F& \multirow{11}{*}{VQE} \\
\cline{2-13}
 & 17& 0.10&0.20 &0.10  & 0.02& 17 & 0.10 & 0.20 & 0.10 & 0.02 & ibmq\_johannesburg & F& \\
\cline{2-13}
 &16 &0.10 & 0.10&  0.10&0.0 & 17 & 0.10 & 0.10 & 0.10 & 0.0 & ibmq\_montreal & F& \\
\cline{2-13}
 &13 &0.10 &0.10 & 0.10 &0.0 & 13 & 0.10 & 0.10 & 0.10 & 0.0 &ibmq\_cambridge & F& \\
\cline{2-13}
 &13 &0.11 & 0.20&0.10  &0.04& 19 & 0.10 & 0.20 & 0.10 & 0.02 & ibmq\_rochester & F& \\
\cline{2-13}
 &9 &0.10 &0.10 & 0.10 & 0.0& 14 & 0.10 & 0.10 & 0.10 & 0.0 &  ibmq\_boeblingen & F&  \\
\cline{2-13}
 &13 &0.10 &0.10 &0.10  & 0.0
& 11 & 0.10 & 0.10 & 0.10 & 0.0 & ibmq\_sydney& F &  \\
\cline{2-13}
 &12 &0.10 &0.10 &0.10  &0.10 
& 12 & 0.10 & 0.10 & 0.10 & 0.0 & ibmq\_toronto & F&  \\
\cline{2-13}
 &10 &0.10 &0.10 & 0.10 &0.0 & 12 & 0.10 & 0.10 & 0.10 & 0.0 &  ibmq\_manhattan & F&  \\
 \cline{2-13}
 &11 & 0.11& 0.20& 0.10 &0.03 & 11 & 0.11 & 0.20 & 0.10 & 0.03 &  ibmq\_sydney& T &  \\
\cline{2-13}
 & 10&0.10 & 0.10& 0.10 &0.0 & 9 & 0.10 & 0.10 & 0.10 & 0.0 &  ibmq\_toronto & T&  \\
\cline{2-14}
 &0 &- &- & - &- & 7 & 0.10 & 0.10 & 0.10 & 0.0 & ibmq\_qasm-simulator & T & \multirow{3}{*}{QAOA}  \\
 \cline{2-13}
 &13 &0.10& 0.10& 0.10 &0.0 &  -&- &- &  &-  & ibmq\_sydney & T &  \\
\cline{2-13}
 &13 &0.10 &0.10 &0.10  &0.0 & 11 & 0.10 & 0.10 & 0.10 & 0.0 &  ibmq\_toronto & T &  \\
\hline
\multirow{9}{*}{5} &8 &28.05 & 48.63&0.10 &18.49 & 29 & 94.24 & 94.24 & 94.24 & 0.0 &  ibmq\_qasm-simulator& F & \multirow{9}{*}{VQE}\\
\cline{2-13}
 &0 &- &- & - &- & 0 & - & - & - & - & ibmq\_montreal& F& \\
\cline{2-13}
 &0 &- &- & - & -& 0 & - & - & - & - & ibmq\_cambridge& F& \\
\cline{2-13}
 &1 &0.10 &0.10 & 0.10 &0.0 & 0 & - & - & - & - & ibmq\_rochester & F&\\
 \cline{2-13}
 &1 & 0.10& 0.10&0.10  &0.0 & 0 & - & - & - & - & ibmq\_sydney &  F&\\
\cline{2-13}
 & 1&0.10 & 0.10& 0.10 &0.0 & 2 & 0.10 & 0.10 & 0.10 & 0.0 & ibmq\_toronto & F& \\
\cline{2-13}
 &1 & 0.10& 0.10& 0.10 &0.0 & 1 & 0.10 & 0.10 & 0.10 & 0.0 & ibmq\_manhattan & F& \\
\cline{2-13}
 &1 &0.10 &0.10 &0.10  &0.0 & 0 & - & - & - & - & ibmq\_sydney & T &  \\
\cline{2-13}
 & 0& -&- &-  &-& 1 & 0.10 & 0.10 & 0.10 & 0.0 & ibmq\_toronto & T &  \\ 
\hline
\multirow{2}{*}{6} & 0& -& -& - &- & 1 & 0.10 & 0.10 & 0.10 & 0.0 & ibmq\_rochester & F & \multirow{2}{*}{VQE}\\
\cline{2-13}
 &0 &- &- & - & -& 0 & - & - & - & - & ibmq\_manhattan & F & \\
 \hline 
7 & 0& -& -& - &- & 0 & - & - & - & - &ibmq\_manhattan & F & VQE \\
\hline\bottomrule
\end{tabular}
}
\end{center}
\begin{tablenotes}\footnotesize
\item[$\dagger$] \# feas. - number of the 30 trials that are feasible
\item [] CR - conditional reset (T for True and F for False)
\end{tablenotes}
\end{threeparttable}
\end{table*}

\section{Conclusion}\label{sec:conclusion}

Many important problems facing industry and academia alike often take form in the class of intractable combinatorial problems. Due to this intractability, significant research in developing classical optimisation approaches has been undertaken. However, with the emergence of quantum computing and its potential in removing this classical limitation, investigating available quantum devices' efficacy on well-posed COP is an important area of current and future research.

With the access to IBM's NISQ devices becoming increasingly available, it is necessary to investigate how well they perform in solving instances of well-known COP problems and their comparative performance to well-accepted classical solutions in the literature.

The results evidence the current limitations of NISQ devices. The selected classical optimisation techniques outperformed the NISQ devices in computational time, SR, feasibility and uncertainty percentage. Both the VQE and QAOA had similar performance with respect to accuracy; however, VQE yielded better computational performance and capacity in solving larger problem size instances. Furthermore, the recently added conditional reset feature was tested but showed no significant improvement in any reported metrics. Importantly, however, the computational results presented herein extend and agree with other findings in the literature \cite{srinivasan2018efficient,ajagekar2019quantum,chieza2020computational}.

Quantum technology is still in its infancy, and these NISQ devices are no exception. Therefore, while the findings of this work show that existing classical devices significantly outperform the performance of current NISQ devices, their performance could prospectively improve as quantum technology evolves.
For example, recent formulations, such as QUBO and ADMM, and the promise of IBM devices with higher-performing processors with more qubits provide hope for higher quality solutions for COP and should be pursued in future work.

\section*{Acknowledgement}
The authors of this research paper acknowledge the \href{http://www.wits.ac.za}{University of the Witwatersrand, Johannesburg} contribution through its support and Quantum Computing resources, which made this research possible. We acknowledge the use of IBM Quantum services for this work. The views expressed are those of the authors, and do not reflect the official policy or position of IBM or the IBM Quantum team.

\section*{Declarations}

\subsubsection*{Funding}

Not applicable.

\subsubsection*{Conflict of Interest/Competing interests}

The authors declare that they have no conflict of interest.

\subsubsection*{Availability of Data and Material}

The open-source data libraries used are referenced in the paper.

\subsubsection*{Code Availability}

The code used to conduct this research has been made available on GitHub. Details to the repository are noted in the paper.

\bibliographystyle{spbasic}      
\bibliography{reference.bib}   

%
%

\end{document}